\documentclass[prl,superscriptaddress,floatfix,twocolumn,showpacs,amsfonts,amsmath,amssymb]{revtex4}
\usepackage{graphicx} 
\usepackage{dcolumn} 
\usepackage{bm} 

\begin{document}

\title{ Heavy quark spin structure in $Z_b$ resonances}

\author{A.E. Bondar}
\affiliation{Budker Institute of Nuclear Physics, 630090
Novosibirsk, Russia}
\affiliation{Novosibirsk State University,
630090 Novosibirsk, Russia}
\author{A. Garmash}
\affiliation{Budker Institute of Nuclear Physics, 630090
Novosibirsk, Russia}
\affiliation{Novosibirsk State University,
630090 Novosibirsk, Russia}
\author{A.I. Milstein}
\affiliation{Budker Institute of Nuclear Physics, 630090
Novosibirsk, Russia}
\affiliation{Novosibirsk State University,
630090 Novosibirsk, Russia}
\author{R. Mizuk}
\affiliation{Institute of Theoretical and Experimental Physics, Moscow, Russia}
\author{M.B.Voloshin}
\affiliation{Institute of Theoretical and Experimental Physics, Moscow, Russia}
\affiliation{William I. Fine Theoretical Physics Institute,
University of Minnesota,
Minneapolis,  USA}

\begin{abstract}
We discuss the heavy quark spin structure of the recently observed
`twin' resonances $Z_b(10610)$ and $Z_b(10650)$ assuming that these
are mostly of a `molecular' type, i.e. that their internal dynamics
is dominated by the coupling to meson pairs $B^* \bar B - B {\bar
B}^*$ and $B^* {\bar B}^*$. We find that the state of the $b \bar b$
pair within the $Z_b(10610)$ and $Z_b(10650)$ resonances is a
mixture of a spin-triplet and a spin-singlet of equal amplitude and
with the phase orthogonal between the two resonances. Such a
structure gives rise to specific relations between observable
amplitudes that are in agreement with the data obtained recently by
Belle. We also briefly discuss possible properties of the
isotopically singlet counterparts of the newly found resonances, and
also of their $C$ ($G$) parity opposites that likely exist in the
same mass range near the open $B$ flavor threshold.
\end{abstract}


\maketitle

Very recently the isotriplet resonances $Z_b(10610)$ and
$Z_b(10650)$ were discovered in the processes $\Upsilon(5S)\to
\pi\pi h_b(kP)$, and $\Upsilon(5S)\to \pi\pi \Upsilon(nS)$,
Ref.~\cite{Belle}. Here and bellow $n=1,2,3$ and $k=1,2$. For
simplicity we refer to $Z_b(10610)$ and $Z_b(10650)$ as $Z_b$ and
$Z_b'$. Data analisis has shown that these processes go mainly as
cascades, e.g., $\Upsilon(5S)\to Z_b \pi\to h_b\pi\pi$. The new
bottomonium-type resonances apparently have quantum numbers
$I^G(J^P)=1^+(1^+)$, so that their electrically neutral isotopic
states with $I_3=0$ should have $J^{PC}=1^{+-}$. It turns out that
the process with $\Upsilon(nS)$ in the final state has almost the
same  probability as those with $h_b$. At first glance this fact
looks quite astonishing. Assuming that $b$ and $\bar b$ quarks are
in a triplet spin state in $\Upsilon(5S)$ and $\Upsilon(nS)$, and
they are in a singlet spin state in $h_b$, one may expect that a
process with spin flip should be suppressed in comparison with that
without spin flip because of the large mass of the $b$ quark. In
this Letter we suggest an explanation of the puzzle. First of all we
note that the masses of the newly found states $Z_b$ and $Z_b'$ are
close to  the respective thresholds of the open $B$ flavor channels
$B^* {\bar B}$ and $B^* {\bar B}^*$. Therefore it is natural to
suggest that the new resonances have a `molecular' type structure of
$B(B^*)$ meson pairs. Namely, the states with the quantum numbers of
$Z_b$ and $Z_b'$ can be realized as $S$-wave $B^* {\bar B}$ and $B^*
{\bar B}^*$ meson pairs, respectively. A possible `molecular' type
structure in the charmonium family was initially discussed in Ref.
~\cite{vo}.

The  mass differences between the charged and neutral $B(B^*)$
mesons are negligible so that, unlike at the charm threshold, the
isotopic symmetry should be well applicable to bottomonium-like
multiquark states. Thefore, we suggest that at long distances, $r
\gg \Lambda_{QCD}^{-1}$, the wave functions  of the $Z_b$ and $Z_b'$
resonances are that of an $S$-wave meson pair in the
$I^G(J^P)=1^+(1^+)$ state, namely, $B^*{\bar B} -  B{\bar B}^*$ for
the $Z_b$ and $B^* {\bar B}^*$ for the $Z_b'$. At shorter distances,
$r \sim \Lambda_{QCD}^{-1}$, the mesons overlap and form a system
containing the heavy quark pair and a light component of quarks and
gluons with the quantum numbers of an isotopic triplet.

In the limit $m_b>>\Lambda_{QCD}$, the spin degrees of freedom  of
$b$ quark in the wave functions $\Psi$ of $B$ ($B^*$) mesons can be
separated from other degrees of freedom. Thus we treat a hypefine
interaction in $B$-meson as a perturbation. As a result, the wave
function $\Psi$ can be written as a direct product ${\bar\psi}_{\bar
q}\otimes\chi_b $, where  spinor $\chi_b$ describes the spin state
of $b$-quark and $\psi_{\bar q}$ describes the wave function of the
bound light antiquark ${\bar q}$ and spinless $b$-quark.  The total
angular momentum $j$ corresponding to the wave function $\psi$ is
fixed in the ground state $B(B^*)$: $j=1/2$, and the rules of
constracting the wave function $\Psi$ are the same as in the
nonrelativistic quark model.The precision of this picture is
determined by the ratio $\Lambda_{QCD}/m_b = O(0.1)$ and the
expected corrections should be at the level of about 10\%.

For $B$ meson we have $\Psi_B={\bar\psi}_{\bar q}\chi_b $, and for
$B^*$ meson we have ${\vec\Psi}_{B^*}={\bar\psi}_{\bar q}{\vec
\sigma}\chi_b $, where ${\vec \sigma}$ are the Pauli matrices.  Then the
$S$- state of the heavy meson pairs with the appropriate quantum
numbers $I^G(J^P)=1^+(1^+)$ is $B^* {\bar B}^*$:
\begin{eqnarray}\label{bsbs}
&&i \, \epsilon_{i j k} \, ({\bar\chi}_{\bar b}\sigma^j\psi_q) ({\bar\psi}_{\bar Q} \sigma^k \chi_b)\nonumber\\
&& =({\bar\chi}_{\bar b}\chi_b) ({\bar\psi}_{\bar Q} \sigma^i\psi_q) -({\bar\chi}_{\bar b}\sigma^i\chi_b) ({\bar\psi}_{\bar Q}\psi_q)\nonumber\\
&& \sim 0^-_{{\bar b}b} \otimes 1^-_{{\bar Q}q}-  1^-_{{\bar b}b} \otimes 0^-_{{\bar Q}q} \,,
\end{eqnarray}
for $B^* {\bar B}^*$, and
\begin{eqnarray}\label{bsb}
&&({\bar\chi}_{\bar b}\sigma^i\psi_q) ({\bar\psi}_{\bar Q}\chi_b)+({\bar\chi}_{\bar b}\psi_q) ({\bar\psi}_{\bar Q}\sigma^i\chi_b)\nonumber\\
&& =-({\bar\chi}_{\bar b}\chi_b) ({\bar\psi}_{\bar Q} \sigma^i\psi_q) -({\bar\chi}_{\bar b}\sigma^i\chi_b) ({\bar\psi}_{\bar Q}\psi_q)\nonumber\\
&& \sim 0^-_{{\bar b}b} \otimes 1^-_{{\bar Q}q}+1^-_{{\bar b}b} \otimes 0^-_{{\bar Q}q}             \,,
\end{eqnarray}
$B^* (\bar B) - B {\bar B}^*$. Here we used the Fierz transforms,
$0^-$ and $1^-$  stand for para- and ortho- states with the negative
parity. Clearly the relations (\ref{bsbs}) and (\ref{bsb}) refer
only to the spin variables of the quarks. These relations describe
the perfect mixtures of the two possible states corresponding to the
para- and ortho- spin states of the $b \bar b$ pair. We thus
conclude that, if the $Z_b'$ and $Z_b$ peaks are determined by a
molecular dynamics of the meson pairs, their heavy quark spin
structure should be the same as of the pairs, i.e.
\begin{eqnarray}\label{pm}
 | Z_b' \rangle&=& \frac{1}{\sqrt{2}}\left( 0^-_{{\bar b}b} \otimes 1^-_{{\bar Q}q}-  1^-_{{\bar b}b} \otimes 0^-_{{\bar Q}q}\right)\,,\nonumber\\
| Z_b \rangle &=& \frac{1}{\sqrt{2}}\left( 0^-_{{\bar b}b} \otimes 1^-_{{\bar Q}q}+  1^-_{{\bar b}b} \otimes 0^-_{{\bar Q}q}\right)\,  .
\end{eqnarray}

Since the masses of  $Z_b$ and $Z_b'$ are very close to sum of the
$B$ and $\bar B^*$ masses and $B^*$ and $\bar B^*$ masses,
respectively, the mixture of states in  Eq.(\ref{pm}) is small. In
this picture the mass splitting between the peaks should be equal to
that between the $B$ and $B^*$ mesons: $M(Z_b')-M(Z_b) = M(B^*)-M(B)
\approx 46\,$MeV with an expected correction of order
$\Lambda_{QCD}^2/m_b = O(1 \div 5)$\,MeV. The spin structure
described by Eq.(\ref{pm}) also leads to an important and
experimentally testable conclusion that  the resonances $Z_b$ and
$Z_b'$ should have the same width. Indeed, in the large $m_b$ limit
all the ortho- and para- states of the $b \bar b$ pair are
degenerate, so that the antisymmetric and the symmetric
superposition of the spin states in Eq.(\ref{pm}) decay into
degenerate (and orthogonal) states with lower mass, so that the
total decay rates of the discussed resonances should be almost
equal: $\Gamma(Z_b)=\Gamma(Z_b')$. In particular, this also implies
that in our approximation the decays of the type $Z_b' \to B^* \bar
B$ are forbidden by the heavy quark spin symmetry, in spite of being
perfectly allowed by the overall quantum numbers and the kinematics.
In other words, the heavy quark spin wave function in the $Z_b'$ is
orthogonal to that in the $Z_b$ state.

The maximal ortho-para mixing of the heavy quarks in the $Z_b$ and
$Z_b'$ resonances described by Eq.(\ref{pm}) immediately implies
that these resonances have coupling of comparable strength to
channels with states of ortho- and para- bottomonium. Furthermore,
for each specific channel the absolute value of the coupling is the
same for $Z_b$ and $Z_b'$. However the relative phase of the
coupling of these resonances to the ortho- bottomonium is opposite
to that for the para- bottomonium. In particular the coupling of
these resonances to the channels $\Upsilon(n S) \, \pi$ and $h_b(kP)
\, \pi$ can readily be found (up to an overall normalization) as
\begin{equation}\label{uhz} E_\pi \, {\vec \Upsilon} \cdot (\vec Z_b
- \vec Z_b')~, ~~~~(\vec p_\pi \times \vec h_b) \cdot (\vec Z_b +
\vec Z_b')~,
\end{equation}
with $\vec Z_b', \, \vec \Upsilon$ and $\vec h_b$ standing for the
polarization amplitude of the corresponding spin one state, and
$E_\pi$  and $\vec p_\pi$ being the pion energy and momentum.  The
amplitudes described by Eq.(\ref{uhz}) can be directly applied to
the resonance part of the amplitudes of the observed transitions
$\Upsilon(5S) \to \Upsilon(nS) \, \pi^+ \, \pi^-$ and $\Upsilon(5S)
\to h_b(kP) \, \pi^+ \, \pi^-$. We have
\begin{eqnarray}
\label{uupp}
&&A(\Upsilon(5S) \to \Upsilon(nS) \, \pi^+ \, \pi^-) = A_\Upsilon ^{\rm nr}
+C_\Upsilon \, (\vec \Upsilon_5 \cdot \vec \Upsilon) \nonumber\\
&&\times \,  E_+ \, E_- \Bigg( {1 \over E - E_+ + { i \over 2} \, \Gamma  }
 + {1 \over E' - E_+ +{ i \over 2} \, \Gamma '  }\nonumber\\
&&+ {1 \over E - E_- + { i \over 2} \, \Gamma  } + {1 \over E' - E_- + { i \over 2} \, \Gamma ' } \Bigg)\,,
\end{eqnarray}
and
\begin{eqnarray}\label{uhpp}
&&A(\Upsilon(5S) \to h_b(kP) \, \pi^+ \, \pi^-) =  A_h ^{\rm nr}+C_h  \nonumber \\
&&\times\Bigg\{ \left [ \vec \Upsilon_5 \cdot (\vec p_- \times \vec h_b) \right ] \,  E_+  \nonumber \\
&&\times  \left ( {1 \over E - E_+ + { i \over 2} \, \Gamma  } - {1 \over E' - E_+ + { i \over 2} \, \Gamma' } \right ) \nonumber \\
&&+ \left [ \vec \Upsilon_5 \cdot (\vec p_+ \times \vec h_b)  \right ] \,  E_- \nonumber \\
&&\times\left ( {1 \over E - E_- + { i \over 2} \, \Gamma  } - {1 \over E' - E_- +{ i \over 2} \, \Gamma '  }  \right ) \Bigg\}\,,
\end{eqnarray}
where $E_+$ and $\vec p_+$  ($E_-$ and $\vec p_-$) stand for the
energy and momentum of the positive (negative) pion, the parameters
$E$ and $\Gamma$ ($E'$ and $\Gamma'$) are those of the $Z_b$
($Z_b'$) resonance with $E=M[\Upsilon(5S)]-M[Z_b]$ and $E'=
M[\Upsilon(5S)]-M[Z_b']$, the vector $\vec \Upsilon_5$ is the
polarization amplitude of the initial $\Upsilon(5S)$ resonance and
$\vec \Upsilon$ and $\vec h_b$ are the same for respectively the
final $\Upsilon(nS)$ and $h_b(kP)$ resonances.  Furthermore, the
coefficients $C_\Upsilon$ and $C_h$ are constants, and, finally,
$A_{\Upsilon,h} ^{\rm nr}$ are the corresponding nonresonant
amplitudes. The latter amplitudes  generally depend on the
polarizations and the kinematical variables~\cite{bc,mv75} and can
be studied in much the same way as for other similar two-pion
transitions between heavy quarkonium states. It can be stated
however that the nonresonant amplitude $A_h ^{\rm nr}$ should be
heavily suppressed due to the heavy quark spin symmetry and an
absence of enhancing factors~\cite{mv86}. In  Eqs.(\ref{uupp}) and
(\ref{uhpp}) we take into account two isotopic resonant branches,
through the $Z_b^+$ ($Z_b'^+$) and through $Z_b^-$ ($Z_b'^-$).

Clearly, the equations (\ref{uupp}) and (\ref{uhpp}) describe two
different patterns of the interference between the $Z_b$ and $Z_b'$
resonances in the two considered transitions. In the process
$\Upsilon(5S) \to \Upsilon(nS) \, \pi^+ \, \pi^-$, the interference
is destructive when the energy of one of the pions lies  between the
positions of the resonances, $E'$ and $E$, and the interference is
constructive when both energies lie outside of the `twin' resonance
band on the Dalitz plot. In the transition $\Upsilon(5S) \to h_b(kP)
\, \pi^+ \, \pi^-$, the pattern is exactly the opposite: the
interference is constructive inside the resonance band and is
destructive outside the band. In fact, the probability of the latter
transition outside the `twin' resonance band on the Dalitz plot
should be very small due to the mentioned suppression of the
nonresonant amplitude $A_h ^{\rm nr}$. Such picture is fully
supported by the experimental results Ref.~\cite{Belle}.

Based on the considerations presented here, one can  expect
existence of hadronic transitions from the $Z_b$ and $Z_b'$
resonances to other ortho- and para- bottomonium states. In
particular, the transitions $Z_b \to \eta_b \, \rho$ are of an
$S$-wave type and a significant rate is possible for this process.
Following the same consideration as presented above, the resonance
amplitudes of the cascade $\Upsilon(5S)  \to \eta_b \, \rho \pi$
should have opposite sign between the $Z_b$ and $Z_b'$ Breit-Wigner
factors. The processes of the type $Z_b(Z_b')\to \chi_b(1P) \, \pi
\, \pi$ are also kinematically possible, but could be suppressed
because  two pions have to be in the $I^G=1^+$ state and the $\rho$
peak is beyond the kinematical region.  The finding of the $Z_b$ and
$Z_b'$ resonances may call for revisiting the analyses of the
previously known processes, such as the transitions $\Upsilon(3S,4S)
\to \Upsilon(1S,2S) \, \pi \, \pi$ as well as the decay
$\Upsilon(3S) \to h_b(1P) \, \pi \, \pi$ for which a significant
upper bound has become available recently~\cite{babar}. A
contribution of an isovector bottomonium-like resonance in the decay
$\Upsilon(3S) \to \Upsilon(1S) \, \pi \, \pi$ was in fact discussed
some time ago~\cite{mv83,absz}.

The existence or non-existence of `molecular'  bottomonium-like
resonances depends on details of a yet unknown dynamics. However,
the very existence of the $Z_b$ and $Z_b'$ resonances necessarily
implies that additional isovector peaks also exist. Indeed, the
resonance properties are determined by the interaction of the
quasiparticles, which are the bound states of light quark and
spinless $b$ antiquark, while the spin of the $b \bar b$ pair plays
only a `classificational' role. In particular the emergence of the
$Z_b(Z_b')$ resonances can be due to a near threshold singularity in
either the $0^-_{{\bar Q}q}$ or the $1^-_{{\bar Q}q}$ state (or
both) in the $I=1$ channel. The total width of the $Z_b$ and $Z_b'$
in the range $15 \div 20$\,MeV makes the distinction between the
specific type of the singularity, a shallow bound state, a virtual
state, or a resonance, rather moot. In either of these cases there
should exist a pair of isotriplet singularities at the respective
$B^*\bar B^*$ and $B\bar B$ thresholds with spin 0 and the $G$
parity opposite to that of $Z_b(Z_b')$, i.e. with
$I^G(J^P)=1^-(0^+)$. This is because the threshold $S$-wave states
of $B^*\bar B^*$ and $B\bar B$ pairs with such overall quantum
numbers are expressed as mixed combinations of the light and heavy
quark spin states:
\begin{eqnarray}\label{bsbs0}
&&\left. \left ( B^*\bar B^* \right ) \right |_{1^-(0^+)} \sim \frac{ ({\bar\chi}_{\bar b}{\bm \sigma}\psi_q) ({\bar\psi}_{\bar Q}{\bm \sigma} \chi_b)}{\sqrt{3}}\nonumber\\
&& =\frac{\sqrt{3}}{2}({\bar\chi}_{\bar b}\chi_b) ({\bar\psi}_{\bar Q}\psi_q) -\frac{1}{2}\frac{({\bar\chi}_{\bar b}{\bm\sigma}\chi_b) ({\bar\psi}_{\bar Q}{\bm\sigma}\psi_q)}{\sqrt{3}}\nonumber\\
&& \sim \frac{\sqrt{3}}{2}0^-_{{\bar b}b} \otimes 0^-_{{\bar Q}q}-
\frac{1}{2} 1^-_{{\bar b}b} \otimes 1^-_{{\bar Q}q} \,,
\end{eqnarray}
and
\begin{eqnarray}\label{bb0}
&&\left. \left ( B \bar B \right ) \right |_{1^-(0^+)} \sim ({\bar\chi}_{\bar b}\psi_q) ({\bar\psi}_{\bar Q}\chi_b)\nonumber\\
&& =\frac{1}{2}({\bar\chi}_{\bar b}\chi_b) ({\bar\psi}_{\bar Q}\psi_q) +\frac{\sqrt{3}}{2}\frac{({\bar\chi}_{\bar b}{\bm\sigma}\chi_b) ({\bar\psi}_{\bar Q}{\bm\sigma}\psi_q)}{\sqrt{3}}\nonumber\\
&& \sim \frac{1}{2}0^-_{{\bar b}b} \otimes 0^-_{{\bar
Q}q}+\frac{\sqrt{3}}{2}1^-_{{\bar b}b} \otimes 1^-_{{\bar Q}q} \,.
\end{eqnarray}
Thus, the mixing angle between para- and ortho- spin states of the
$b \bar b$ pair in this case is $\pi/6$ which can be readily checked
experimentally. Clearly, such resonances form another 'twin' pair
and couple to both ortho- and para- bottomonium and can thus decay
to e.g. $\Upsilon \, \rho$ as  well as to $\eta_b \, \pi$.

Additionally, if a threshold singularity  $1^-_{{\bar Q}q}$ state
contributes to the $Z_b(Z_b')$ resonances, its combination with
$1^-_{{\bar b}b}$ state of the heavy quark pair should also produce
an $I^G(J^P)=1^-(1^+)$ state at the $B^* \bar B$ threshold (an
isovector bottomonium analog of X(3872)Ref.~\cite{X3872}), and
isospin triplet at the $B^* \bar B^*$ threshold with spin 2:
$I^G(J^P)=1^-(2^+)$. These latter states couple to
ortho-bottomonium, e.g., to the channel $\Upsilon \, \rho$. Due to
the negative $G$ parity all these expected resonances are not
accessible in single pion transitions from $\Upsilon(5S)$, but their
production can be sought for at a somewhat higher energy above the
$B$ flavor threshold. One can also notice that the isotriplet states
cannot mix with pure bottomonium states, so that they are unlikely
to be produced at Tevatron and/or LHC at a detectable rate.

At this point the `molecular' interaction in the isoscalar channel
is not known. However, based on the existence of the $X(3872)$ state
in the charmonium family, one may expect an existence of $I=0$
counterparts of the isotriplet $Z_b$ and $Z_b'$ in the same mass
region near the $B^* \bar B$ and $B^* \bar B^*$ thresholds. Such
states, $Y_b$ and $Y_b'$ have the quantum numbers $J^{PC}=1^{+-}$
and can mix with $^1P_1$ states of bottomonium,  $h_b(kP)$. Such
mixing can generally tilt the completely mixed orto- para- heavy
quark spin structure in the $Y_b(Y_b')$ resonances, and it would be
interesting if this behavior could be studied experimentally. For
the reasons of isospin these resonances are not accessible from the
$\Upsilon(5S)$ by single pion transitions, but could be studied in
the future at higher initial energies in $e^+e^-$ annihilation.
Moreover, the likely presence of the bottomonium $^1P_1$ `core' in
the $Y_b(Y_b')$ states makes it possible that, unlike the
$Z_b(Z_b')$, these states can be produced in hard processes such as
the high-energy $p \bar p$ or $pp$ collisions at the Tevatron and
LHC, similarly to the production of $X(3872)$ at the
Tevatron~\cite{cdf}. The discussed bottomonium-like resonances can
be identified, e.g., by their decay into $\Upsilon(2S) \, \eta$, or
$\Upsilon(1S) \, \eta$, or $\Upsilon(1S) \, \eta'$ which all are
$S$-wave processes and which one would not expect to be suppressed.
Other possibly identifiable in a collider setting decay channels of
$Y_b(Y_b')$ are $\Upsilon(1S) \, \pi \, \pi$ and $\Upsilon(1S) \, K
\, {\bar K}$, including those through the $f_0(980)$ resonance:
$Y_b(Y_b') \to \Upsilon(1S) \, f_0(980)$, although an expectation
for the rate of these latter processes is a somewhat subtle due to
the required orbital momentum of the light mesons. A comparison of
the decay rates to states with and without hidden strangeness could
also shed light on the significance of an admixture in these
resonances of the states of the type $b \bar b s \bar s$. The
$C$-even states $X_b$ of the same type can mix with the $^3P_J$
bottomonium and can similarly be produced in hard collisions. These
resonances can be sought for at the colliders by their decay, e.g.,
into $\Upsilon(1S) \, \omega$.

In summary. We argue that, if the newly found $Z_b$ and $Z_b'$
isovector resonances are states of a `molecular' type in
respectively the channels $B^* \bar B -  B{\bar B}^*$ and $B^*
{\bar B}^*$ with the quantum numbers $I^G(J^P)=1^+(1^+)$, each of
them has to contain (almost) complete mixture of spin-triple and
spin-singlet states of the heavy $b \bar b$ pair. The heavy quark
spin wave functions in the two resonances have to be orthogonal to
each other, as described by Eq.(\ref{pm}). In our approach using a
separation of the $b$ quark spin degrees of freedom in the wave
function of $B$ mesons, based on the large value of the $b$ quark
mass $m_b$, the mass splitting between $Z_b'$ and $Z_b$ should be
the same as between $B^*$ and $B$ mesons:
$M(Z_b')-M(Z_b)=M(B^*)-M(B) \approx 46$\,MeV, and their total decay
widths should be equal to one another: $\Gamma(Z_b')= \Gamma(Z_b)$.
Any deviations from these relations are due to the finite mass of
$b$ quark and should be small. In particular, a kinematically
allowed process $Z_b' \to B^* \bar B$ should be strongly suppressed.
Furthermore, the resonances $Z_b$ and $Z_b'$ should have equal
coupling to specific decay channels with the states of bottomonium.
The relative sign between the couplings of $Z_b$ and $Z_b'$ to such
channels depends on the spin state of the bottomonium, this relative
sign of the coupling to ortho- states is opposite to that in the
coupling to the para- states. Such behavior leads to a specific
interference pattern in the contribution of the discussed resonances
to the observed processes $\Upsilon(5S) \to \Upsilon(nS) \, \pi^+ \,
\pi^-$ and $\Upsilon(5S) \to h_b(kP) \, \pi^+ \, \pi^-$. Finally, we
point out that a similar behavior can be tested in the yet
unobserved, but expected, processes $\Upsilon(5S) \to \eta_b \, \rho
\pi$ and $\Upsilon(5S) \to \chi_b(1P) \, \pi \, \pi \, \pi$. The
coupling of the bottomonium states to the $Z_b$ and $Z_b'$
resonances can also affect the rates and the spectra in hadronic
transitions in bottomonium, such as $\Upsilon(3S) \to \Upsilon(1S)
\, \pi \, \pi$ and/or $\Upsilon(3S) \to h_b(1P) \, \pi \, \pi$. We
also suggest that isospin-singlet resonances $Y_b$ and $Y_b'$ with
the quantum numbers $J^{PC}=1^{+-}$ of a similar `molecular'
structure can mix with the $^1P_1$ states of bottomonium and can
thus be produced in `hard' collisions at the Tevatron and LHC. These
states can be sought for in the high-energy data by their decay
channels $\Upsilon(2S,1S) \, \eta$, or $\Upsilon(1S) \eta'$, or
$\Upsilon(1S) \, \pi \pi \, ({\rm or}~ K \bar K)$, and a possible
isoscalar resonance $X_{b0}$ with $J^{PC}=1^{++}$  can be sought for
by its decay into $\Upsilon(1S) \, \omega$.

This work is supported in part by the RFBR Grant No 09-02-0024 and
the Grant 14.740.11.0082 of Federal Program "Personnel of
Innovational Russia". The work of M.B.V. is supported in part by the
DOE grant DE-FG02-94ER40823.

\end{document}